\documentclass[prd,a4paper]{revtex4}

\usepackage{graphicx}
\usepackage{psfrag}
\usepackage{inputenc}
\usepackage{amsmath}
\usepackage{amssymb}
\usepackage{subfigure}

\inputencoding{latin9}

\newcommand{\nc}{\newcommand}

\def\lsim{\; \raise0.3ex\hbox{$<$\kern-0.75em
      \raise-1.1ex\hbox{$\sim$}}\; }
\def\gsim{\; \raise0.3ex\hbox{$>$\kern-0.75em
      \raise-1.1\textmd{}ex\hbox{$\sim$}}\; }

\nc{\be}[1]{\begin{equation}\mbox{$\label{#1}$}}
\nc{\bea}[1]{\begin{eqnarray} \mbox{$\label{#1}$}}
\nc{\Section}[2]{\section{#2}\label{#1}}
\nc{\Bibitem}[1]{\bibitem{#1}}
\nc{\Label}[1]{\label{#1}}
\nc{\ie}{{\em i.e. }}
\nc{\eg}{{\em e.g. }}
\nc{\eea}{\end{eqnarray}}
\nc{\ee}{\end{equation}}
\nc{\w}{\omega}
\begin{document}

\title{Constraints on the three-fluid model of curvaton decay}

\author{T. Multam\"aki}
\thanks{tuomul@utu.fi}
\affiliation{Department of Physics,
University of Turku, FIN-20014 Turku, FINLAND}
\author{J. Sainio}
\thanks{jtksai@utu.fi}
\affiliation{Department of Physics,
University of Turku, FIN-20014 Turku, FINLAND}
\author{I. Vilja}
\thanks{vilja@utu.fi}
\affiliation{Department of Physics,
University of Turku, FIN-20014 Turku, FINLAND}
\date{\today}

\begin{abstract}
A three fluid system describing the decay of the curvaton is studied by numerical and analytical 
means. 
We place constraints on the allowed interaction strengths between 
the fluids and initial curvaton density by requiring that the curvaton decays before
nucleosynthesis while nucleosynthesis, radiation-matter equality and decoupling occur at  
correct temperatures. 
We find that with a continuous, time-independent interaction, a small 
initial curvaton density is naturally preferred along with a low reheating temperature. Allowing 
for a time-dependent interaction, this constraint can be relaxed. In both cases, a purely 
adiabatic final state can be generated, but not without fine-tuning. Unlike in the two fluid 
system, the time-dependent interactions are found to have a small effect on the curvature 
perturbation itself due to the different nature of the system. The presence of non-gaussianity 
in the model is discussed.

\end{abstract}

\maketitle

\section{Introduction}

The problem of determining the evolution of large scale perturbations in a background of a 
multi-component fluid system is central in modern day cosmology 
\cite{Kodama:1985bj, Mukhanov:1990me, Wands:2000dp}. 
In such a system, the interactions between the different fluids are important in determining the
evolution of the curvature perturbation \cite{Malik:2004tf}. Examples of interacting fluid 
systems demonstrate the importance of such systems,as  most notably reheating at the end of 
inflation
\cite{Albrecht:1982mp,Den:1984tn,Kripfganz:1985mn,Bastero-Gil:2002xr,Dvali:2003em,Kofman:2003nx} 
and the curvaton scenario 
\cite{Enqvist:2001zp,Lyth:2001nq,Moroi:2002rd,Moroi:2001ct,Lyth:2002my}.

In contrast to any single fluid system, in a multi-component system the total curvature 
perturbation, $\zeta$, generally evolves whenever the non-adiabatic pressure is non-zero, 
\ie when interactions between the fluids exist. Evolution of the primordial large scale 
curvature perturbation can relax the underlying assumptions on the inflationary scenario.
Therefore analysis of multi-component fluid systems may affect our view on
the physical settings. In addition to the curvature scenario 
considered in this paper, natural frameworks for such mechanism exist \eg within a traditional 
multiple inflationary scenario \cite{minf} or a string landscape picture 
\cite{cliff, multiverse}. Whether a given scenario can effectively modify the primordial 
spectrum depends on the exact nature of the system.

Recent cosmic microwave surveys have also brought attention into the concept of non-gaussianity 
\ie how much the spectrum deviates from gaussian distribution. This is especially important in 
multifield models of inflation including the curvaton scenario \cite{Malik:2006pm,Sasaki:2006kq}.

The mechanism how energy is transferred between the fluids can be described by different 
methods, \eg by a constant interaction \cite{Malik:2002jb} or by utilizing the so-called sudden
decay approximation \cite{Lyth:2001nq, Lyth:2002my}. In a recent paper \cite{Multamaki:2006} 
we considered relaxing the assumptions behind these approximations by allowing for
time dependent interactions while evolving the full large-scale perturbation equations.
Such an approach can better model the micro physics behind a particular physical framework by 
allowing one to choose the strength and the time at which the interaction is turned on. 
In contrast, if the interaction between is modeled with a constant interaction term, the fluid 
begins to decay (or interact) when its decay width is of the order of the Hubble rate, 
$\Gamma\sim H$. Physical scenarios relevant to having time (and space) dependent interactions 
include \eg phase transitions, multiple inflation scenarios \cite{minf} and scenarios where 
locally different decay rates of the 
inflaton are generated by spatially varying reheating temperature and couplings
\cite{Matarrese:2003tk, Dvali:2003em, Kofman:2003nx}.

In the present paper we consider the curvaton scenario with time dependent interactions between 
curvaton and other fluids. During inflation the curvaton is a light scalar field 
that does not contribute to the expansion of the universe; after the inflaton field has decayed 
into relativistic particles the curvaton begins to oscillate and to decay into radiation and 
matter. The focus of this article is on this situation: we study how a time-dependent interaction
affects the evolution of the curvature perturbation. We study the physically allowed parameter 
space by utilizing information from known cosmological epochs. Moreover, we calculate the amount
isocurvature in terms of curvaton decay widths. Finally, we also discuss how to describe 
non-gaussianity in the three fluid model and present the non-linearity factor $f_{NL}$ 
\cite{Komatsu:2001rj}.

This paper is organized as follows. In sections II and III we present the governing equations of 
the background and perturbations in the Newtonian gauge, including discussion of non-gaussianity 
as described by the $f_{NL}$ parameter. In section IV we present our results while
the discussion and conclusions are presented in the following section V.

\section{Perturbation equations}

We begin by presenting the equations of motion of the background variables and different density 
perturbations, where we adapt the notations and conventions of 
\cite{Mukhanov:1990me,Malik:2004tf} as well as we consider 
linear scalar perturbations about a spatially flat Friedmann-Robertson-Walker -background in a 
Newtonian gauge:
\be{metric}
-ds^2=g_{\mu\nu}dx^{\mu}dx^{\nu}=(1+2\phi)dt^2-a(t)^2(1-2\psi)\delta_{ij}dx^idx^j,
\ee
where $a$ is the scale factor. We have used units where $8\pi G/3=1$ and $c=1$.

The background evolution is determined by the Einstein's equations, 
$G_{\mu\nu}=8\pi G\, T_{\mu\nu}$, and the continuity equations of individual fluids:
\begin{equation} \label{eq:conti}
\dot{\rho}_{(a)} = -3H(1+\omega_{(a)})\rho_{(a)} + Q_{(a)},
\end{equation}
where $\w_{(a)}=P_{(a)}/\rho_{(a)}$ is the equation of state and $Q_{(a)}$ describes the energy 
transfer between different fluids.

Perturbing the covariant continuity equations, one finds the evolution equations of the energy 
and pressure density perturbations $\delta\rho_{(a)}$ and $\delta P_{(a)}$ on large scales 
\cite{Malik:2004tf}:
\be{fluidpert}
\dot{\delta\rho}_{(a)}+3H(\delta\rho_{(a)}+\delta P_{(a)})-3(\rho_{(a)}+P_{(a)})\dot{\psi} = 
Q_{(a)}\phi + \delta Q_{(a)}.
\ee
In addition, we have the $G^{0}_{0}$ component of perturbed Einstein equations
\begin{equation} \label{psieq}
3H\big(\dot{\psi}+H\phi\big) = -4\pi G\delta\rho,
\end{equation}
where $\delta\rho = \sum_a \delta\rho_{(a)}$. For perfect fluids, $\phi=\psi$, and hence given 
the equations of state, $\w_{(a)}$, and the interactions between the fluids, $Q_{(a)}$, one can 
evolve the individual fluid perturbations along with the metric perturbation $\phi$.

Since gauge dependence is an important issue in perturbation analysis we have used gauge 
invariant perturbations $\xi_{(a)}$, which equal density perturbations $\delta \rho_{(a)}$ in 
uniform curvature gauge $\psi = 0$, \ie
\begin{equation} \label{eq:xi-def}
\begin{aligned}
\xi_{(a)} & = \delta \rho_{(a)} + \rho_{(a)}'\psi = -\rho_{(a)}'\zeta_{(a)},\\
\xi & = \sum_{a} \xi_{(a)},\\
\end{aligned}
\end{equation}
where $\zeta_{(a)}$ is the curvature perturbation in uniform density gauge, \ie 
$\delta \rho_{(a)}=0$ and comma represents derivative with respect to the number of e-folds 
$N) \ln a$. 

The corresponding equations of motion for the gauge invariant quantities are thus
\begin{equation} \label{eq:basic-eq}
\begin{aligned}
\xi_{(a)}' & = -3(1+\omega_{(a)})\xi_{(a)} - 3 \delta P_{int(a)} + \frac{\delta Q_{(a)}}{H}+
\frac{Q'_{(a)} \phi}{H} - \frac{Q_{(a)}}{H}\frac{\xi}{2\rho_{0}}\\
& = -3(1+\omega_{(a)})\xi_{(a)} - 3 \delta P_{int(a)} + \frac{1}{H}\Big(\sum_{c}
\frac{\partial Q_{(a)}}{\partial \rho_{(c)}}\xi_{(c)} + \frac{\partial Q_{(a)}}{\partial N} 
\phi\Big) - \frac{Q_{(a)}}{H}\frac{\xi}{2\rho_{0}},\\
\end{aligned}
\end{equation}
where $\delta P_{int(a)} \equiv \delta P_{(a)} - p_{(a)}'\delta \rho_{(a)}/\rho_{(a)}'$ are the 
internal pressure perturbations.
We have also included the possibility of explicit time dependence of the interaction terms.
The equation of motion of the curvature perturbations can be derived from eqs. 
(\ref{eq:basic-eq}) and (\ref{eq:xi-def}). The result is
\begin{equation} \label{eq:zeta-eq}
\begin{aligned}
\zeta_{(a)}' = & \frac{3\delta P_{int(a)}}{\rho_{(a)}'} - \frac{1}{H\rho_{(a)}'}\delta Q_{int (a)} 
-\frac{H'}{H}\frac{Q_{(a)}}{\rho_{(a)}'}(\zeta - \zeta_{(a)}),\\
\end{aligned}
\end{equation}
where $\delta Q_{int (a)} \equiv \delta Q_{(a)}-Q'_{(a)}\delta \rho_{(a)}/\rho'_{(a)}$.

\section{Evolution equations}

The model in hand is a three fluid system with curvaton, radiation and non-relativistic matter 
fields denoted by subscripts$ \sigma$, $\gamma$ and $m$, respectively. 
Since the curvaton field undergoes coherent oscillations, the curvaton fluid can be safely 
estimated \cite{Ferrer:2004nv} to behave like non-relativistic matter. Thus we have
$\omega_{\sigma}=0$, $\omega_{\gamma}=1/3$ and $\omega_{m}=0$ and the background equations 
(\ref{eq:conti}) can be written in terms of fractional densities 
$\Omega_{a} \equiv \rho_{(a)}/\rho$ for which the equations of motion are \cite{Gupta:2003jc}:
\begin{equation} \label{eq:bgr}
\begin{aligned}
\Omega_{\sigma}' & =\Omega_{\sigma}\Omega_{\gamma}+\frac{Q_{\sigma}}{H\rho} ,\\
\Omega_{\gamma}' & =\Omega_{\gamma}(\Omega_{\gamma}-1)+\frac{Q_{\gamma}}{H\rho} ,\\
\Omega_{m}' & =\Omega_{m}\Omega_{\sigma}+\frac{Q_{m}}{H\rho} ,\\
\Big(\frac{1}{H}\Big)' & = \Big(1+\frac{1}{3}\Omega_{\gamma}\Big)\Big(\frac{1}{H}\Big).\\
\end{aligned}
\end{equation}
From the definition of $\Omega_{a}$ it can be easily seen that 
$\Omega_{\sigma}+\Omega_{\gamma}+\Omega_{m}=1$, which means that one of the 
equations of motion
is redundant. 
The interaction terms read as
\begin{equation}
\begin{aligned}
Q_{\sigma} & = -\Gamma_{\gamma}f_{\gamma}(N)\rho_{\sigma} - \Gamma_{m}f_{m}(N)\rho_{\sigma},\\
Q_{\gamma} & = \Gamma_{\gamma}f_{\gamma}(N)\rho_{\sigma},\\
Q_{m} & = \Gamma_{m}f_{m}(N)\rho_{\sigma},\\
\end{aligned}
\end{equation}
where $\Gamma_{(a)}$ denotes the strength of the interaction and functions $f_{(a)}(N)$ include 
the explicit $N$-fold (time) dependence.

From eq. (\ref{eq:basic-eq}) we can finally derive the equations governing the evolution of the 
perturbations of curvaton-radiation-matter system. The equations for gauge invariant 
perturbations are
\begin{equation} \label{eq:xi-pert}
\begin{aligned}
\xi_{\sigma}' & = -3\xi_{\sigma} - \frac{\Gamma_{\gamma}f_{\gamma}(N) + 
\Gamma_{m}f_{m}(N)}{H}\xi_{\sigma} - \frac{\Gamma_{\gamma}f_{\gamma}'(N) + 
\Gamma_{m}f_{m}'(N)}{H}\rho_{\sigma}\phi - \frac{Q_{\sigma}}{H}\frac{\xi}{2\rho}\\
\xi_{\gamma}' & = -4\xi_{\gamma} + \frac{\Gamma_{\gamma}f_{\gamma}(N)}{H}\xi_{\sigma} + 
\frac{\Gamma_{\gamma}f_{\gamma}'(N)}{H}\rho_{\sigma}\phi - \frac{Q_{\gamma}}{H}\frac{\xi}{2\rho}\\
\xi_{m}' & = -3\xi_{m} + \frac{\Gamma_{m}f_{m}(N)}{H}\xi_{\sigma} + 
\frac{\Gamma_{m}f_{m}'(N)}{H}\rho_{\sigma}\phi - \frac{Q_{m}}{H}\frac{\xi}{2\rho}\\
\phi' & = (\frac{H'}{H}-1)\phi - \frac{\xi}{2 \rho}.\\
\end{aligned}
\end{equation}
Correspondingly, the equations for curvature perturbations $\zeta_{a}$ in the uniform density 
gauge read as
\begin{equation} \label{eq:ze-pert}
\begin{aligned}
\zeta_{\sigma}' & = \frac{\Gamma_{m}f'_{m}(N)\rho_{\sigma}(\psi+\zeta_{\sigma})}{H\rho_{\sigma}'} 
+\frac{H'}{H}\frac{(\Gamma_{\gamma}+\Gamma_{m}f_{m}(N))}{\rho_{\sigma}'}(\zeta-\zeta_{\sigma})\\
\zeta_{\gamma}' & = \frac{\Gamma_{\gamma}\rho_{\sigma}'(\zeta_{\sigma}-
\zeta_{\gamma})}{H\rho_{\gamma}'} -\frac{H'}{H}\frac{\Gamma_{\gamma}\rho_{\sigma}}{\rho_{\gamma}'}
(\zeta-\zeta_{\gamma})\\
\zeta_{m}' & = \frac{\Gamma_{m}}{H\rho_{m}'}\Big[f_{m}(N)\rho'_{\sigma}(\zeta_{m}-\zeta_{\sigma})
+f'_{m}(N)\rho_{\sigma}(\psi+\zeta_{m})\Big]-\frac{H'}{H}
\frac{\Gamma_{m}f_{m}(N)\rho_{\sigma}}{\rho_{m}'}(\zeta-\zeta_{\sigma}).\\
\end{aligned}
\end{equation}
The evolution of the system described by eqs (\ref{eq:bgr}) and (\ref{eq:xi-pert}) is then 
ready to be solved numerically once the initial conditions have been set. 

Because some of the initial conditions lead to non-physical solutions (\eg 
too high reheating temperature or the universe might become matter dominated during 
nucleosynthesis), the numerical analysis has to be performed carefully.
In order to eliminate clearly unphysical scenarios we use physical knowledge from notable 
epochs of cosmology, namely Big Bang Nucleosynthesis (BBN), radiation-matter equality and 
decoupling. By fixing the temperatures at those times we are able to determine constraints for
the energy distribution between different components of the 
system. The physical temperature scales have been set to coincide with values given in 
\cite{Liddle:2000}: nucleosynthesis came about at temperature at least $0.1$ MeV, 
the matter-radiation equality was reached when $T=1.0$ eV and decoupling occurred when $T=0.1$ eV.
Because $\rho_{m},\rho_{\sigma} \propto a^{-3}$ and $\rho_{\gamma} \propto a^{-4}$ we can 
derive a lower limit for the abundances of curvaton, CDM and radiation during nucleosynthesis. 
Straightforward calculation results in the limit
\begin{equation}
\frac{\Omega_{\gamma}}{\Omega_{\sigma}+\Omega_{m}}\Big|_{\textrm{nuc}} \geq 10^{5}.
\end{equation}
This limit alone does not, however, guarantee that the curvaton has decayed before the 
nucleosynthesis. Because even a small contribution of curvaton during nucleosynthesis may 
eventually begin to dominate the system afterwards and therefore lead to a system with 
undesirable physical properties, we require that it decays rapidly enough to 
become subdominant. In practice we compel the beginning of nucleosynthesis to happen after the 
curvaton starts to effectively decay, \ie
\begin{equation}
N_{\Omega_{\sigma}, \textrm{max}} \leq N_{ \textrm{nuc}},
\end{equation}
where $N_{\Omega_{\sigma}, \textrm{max}}$ is the time of maximum curvaton proportion.
Setting the temperature at radiation-matter equality allows us to estimate the initial temperature of the system \ie reheating
temperature by $T_{RH}\approx (\rho_{\gamma,RH}/\rho_{\gamma,eq})^{1/4}T_{eq}$.
More precisely, the reheating temperatures we quote here are lower bounds since the decay of the curvaton adds
to the radiation energy density of the system, but numerical work shows that the effect is small and in practice the initial 
temperature can well be approximated by the above formula.

\subsection{Non-gaussianity}

Single field inflation models usually predict the amount of non-Gaussianity to be too low to be 
detectable by future CMB-surveys. In contrast multiple scalar fields (\eg the curvaton scenario)
can lead to a clearly observable non-Gaussianity being therefore testable in near future. 

The non-Gaussianity parameter $f_{NL}$ \cite{Lyth:2002my,Malik:2006pm} is defined via 
gauge invariant Bardeen potential
\begin{equation}
\Phi=\Phi_{g}+f_{NL}\Phi_{g}^{2},
\end{equation}
where $\Phi_{g}$ is the gaussian part of $\Phi$. The Bardeen potential can be expressed at the 
time of decoupling when the universe is matter dominated as $\Phi=3\zeta/5$. The non-Gaussianity
parameter
$f_{NL}$ is common to express utilizing an additional parameter defined by
\begin{equation} \label{eq:r-eq}
r=\frac{\zeta}{\zeta_{\sigma,0}}\Big|_{\textrm{dec}},
\end{equation}
which tells how effectively the initial curvature perturbation transfers to the total curvature 
perturbation. Thereby, using the definition of the curvature perturbation $\zeta$, 
$r$-parameter, the equation of motion of $\rho_{\sigma}$ and a second order estimate 
$\delta\rho_{\sigma}/\rho_{\sigma}=2\delta\sigma/\sigma+(\delta\sigma/\sigma)^{2}$ 
\cite{Lyth:2002my}, the Bardeen potential can be cast in the form
\begin{equation}
\Phi = \frac{2}{5}\frac{r}{1+\frac{\Gamma_{\sigma}}{3H_{0}}}
\Big(\frac{\delta\sigma}{\sigma}\Big)_{0} +  \frac{5}{4}\frac{1+
\frac{\Gamma_{\sigma}}{3H_{0}}}{r}\Bigg(\frac{2}{5}\frac{r}{1+
\frac{\Gamma_{\sigma}}{3H_{0}}}\Big(\frac{\delta\sigma}{\sigma}\Big)_{0}\Bigg)^{2}
\end{equation}
and hence $f_{NL} \simeq 5/(4r)$. Since during decoupling $\zeta\simeq\zeta_{m}|_{\textrm{Dec}}$ 
and in the three-fluid curvaton model $\zeta_{m}|_{\textrm{Dec}}=\zeta_{\sigma,0}$ 
\cite{Gupta:2003jc}, our value for the transfer parameter is 1 and $f_{NL} = 5/4$. As pointed 
out in \cite{Lyth:2002my} the previous estimate of $f_{NL}$ is valid only when $f_{NL}\gg1$ 
since we are using first-order perturbation theory and we are assuming second-order terms 
$\Phi^{(2)}$ to be at most of order $\Phi_g^2$. Thus the three-fluid model is unable to give any 
limitations on the non-Gaussianity parameter when the linear theory is applied.

This result differs from the non-gaussianity given in \cite{Gupta:2003jc} mainly because our 
$f_{NL}$ is defined at the time of decoupling whereas in \cite{Gupta:2003jc} the time of 
nucleosynthesis is used. Our analysis, however, follows the general formalism presented in 
\cite{Bartolo:2004if} which allows our $f_{NL}$ to be compared to the observable first order 
Sachs-Wolfe effect \cite{Mollerach:1997up}
\begin{equation}
\frac{\Delta T}{T}=\phi+\frac{\delta\rho_{\gamma}}{4\rho_{\gamma}},
\end{equation}
which is evaluated at the last scattering surface and where the lapse function $\phi=-\Phi$ in the Newtonian gauge.

\section{Numerical results}

We have studied the evolution of equations (\ref{eq:bgr}) and (\ref{eq:xi-pert}) in two 
physically distinct scenarios: {\bf A} the curvaton decays promptly into radiation and matter \ie 
$f_{\gamma}(N)=1$, $f_{m}(N)=1$ and {\bf B} curvaton decays first only into radiation 
component and the matter interaction begins when $N = N_{*}$ \ie $f_{\gamma}(N)=1$, 
$f_{m}(N)=\theta(N-N_{*})$. In both scenarios we have systematically scanned the parameter space 
in order to identity physically acceptable parameter values.

\subsection{Continuous interactions, $f_{\gamma}(N)=1$, $f_{m}(N)=1$}

\begin{figure}[tbh]

\subfigure[]{\label{fig1a}\includegraphics[width=0.45\columnwidth]{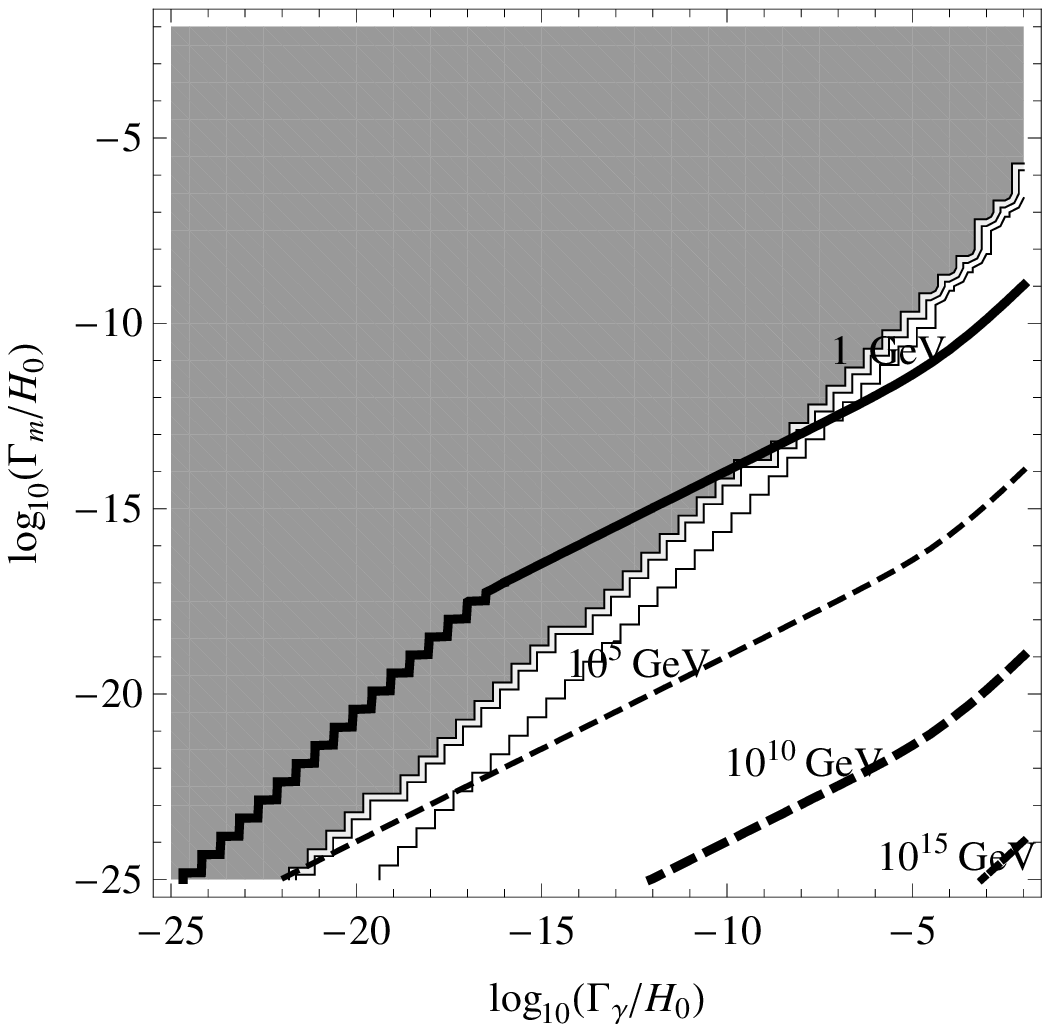}}
\qquad
\subfigure[]{\label{fig1b}\includegraphics[width=0.45\columnwidth]{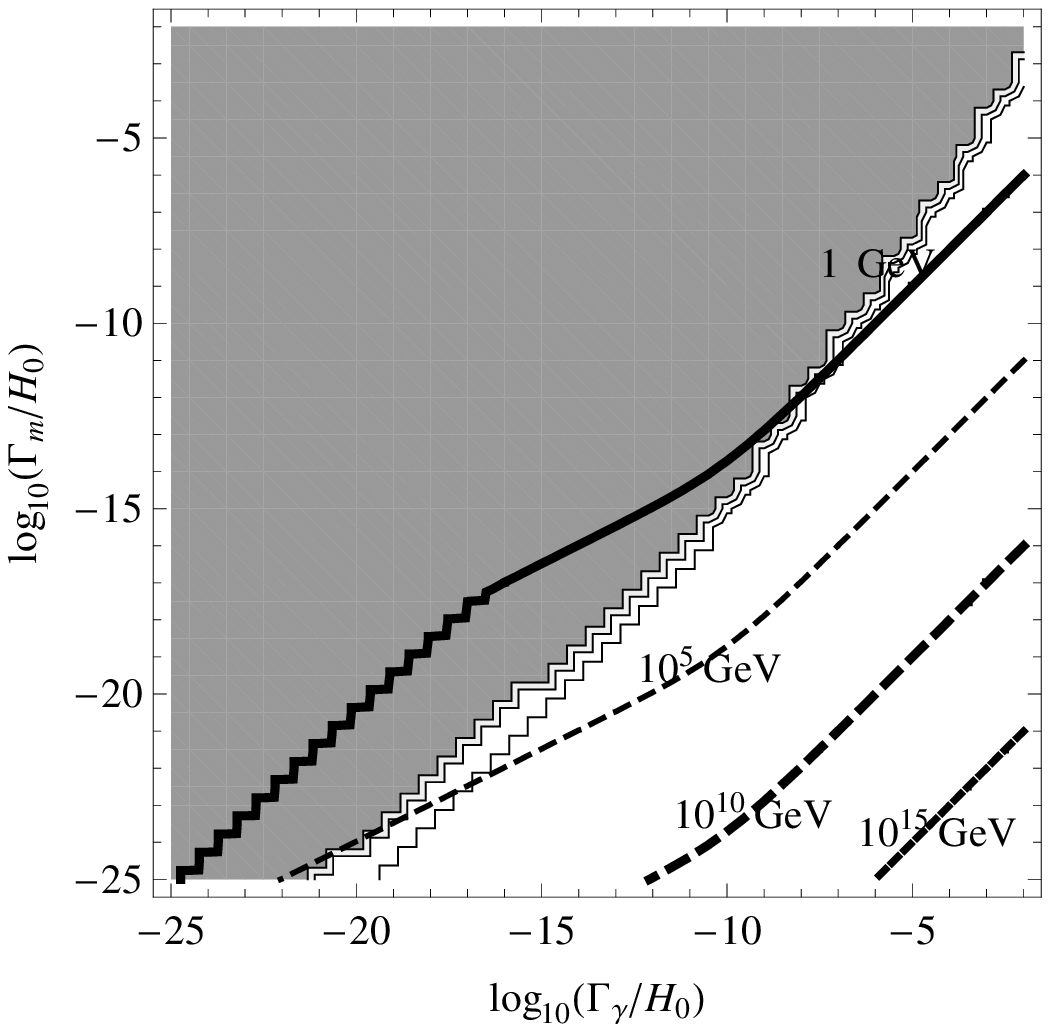}}
\subfigure[]{\label{fig1c}\includegraphics[width=0.45\columnwidth]{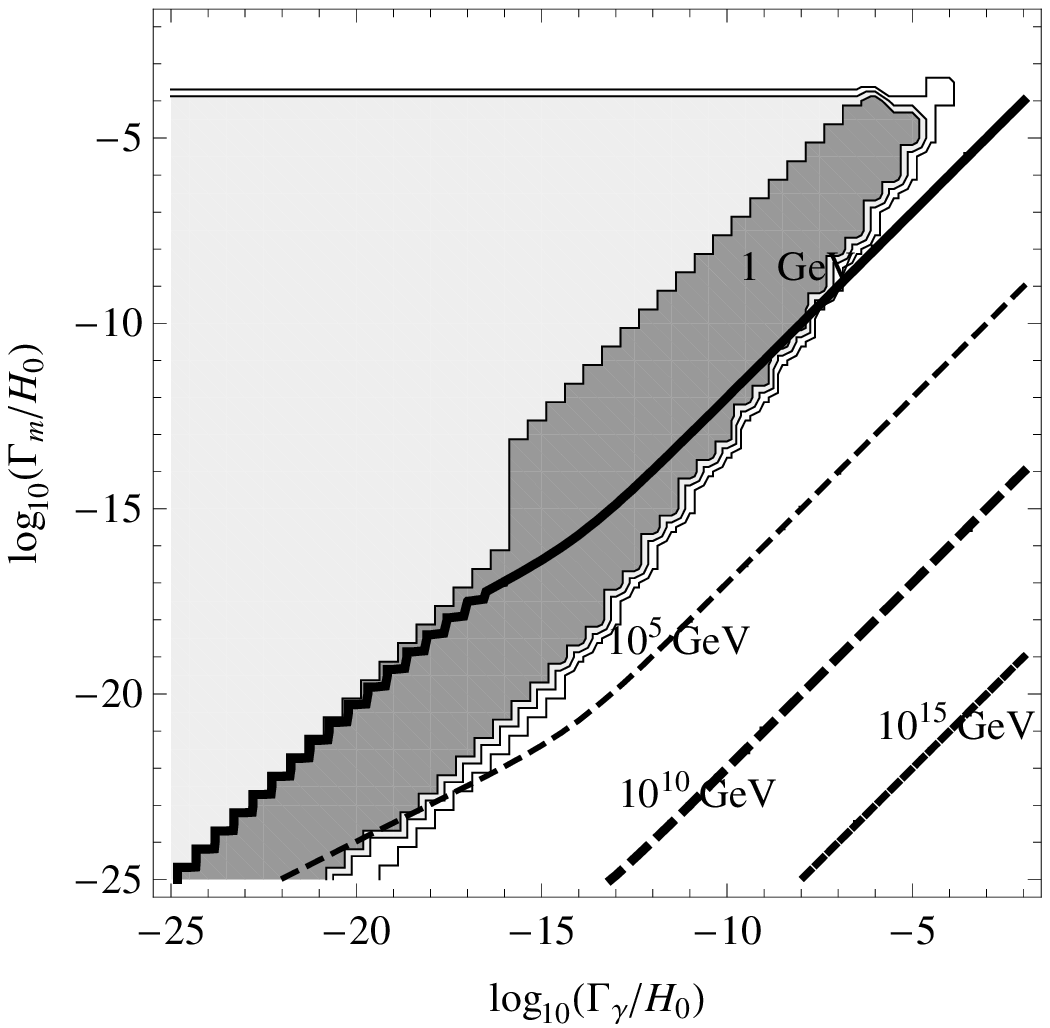}}
\qquad
\subfigure[]{\label{fig1d}\includegraphics[width=0.45\columnwidth]{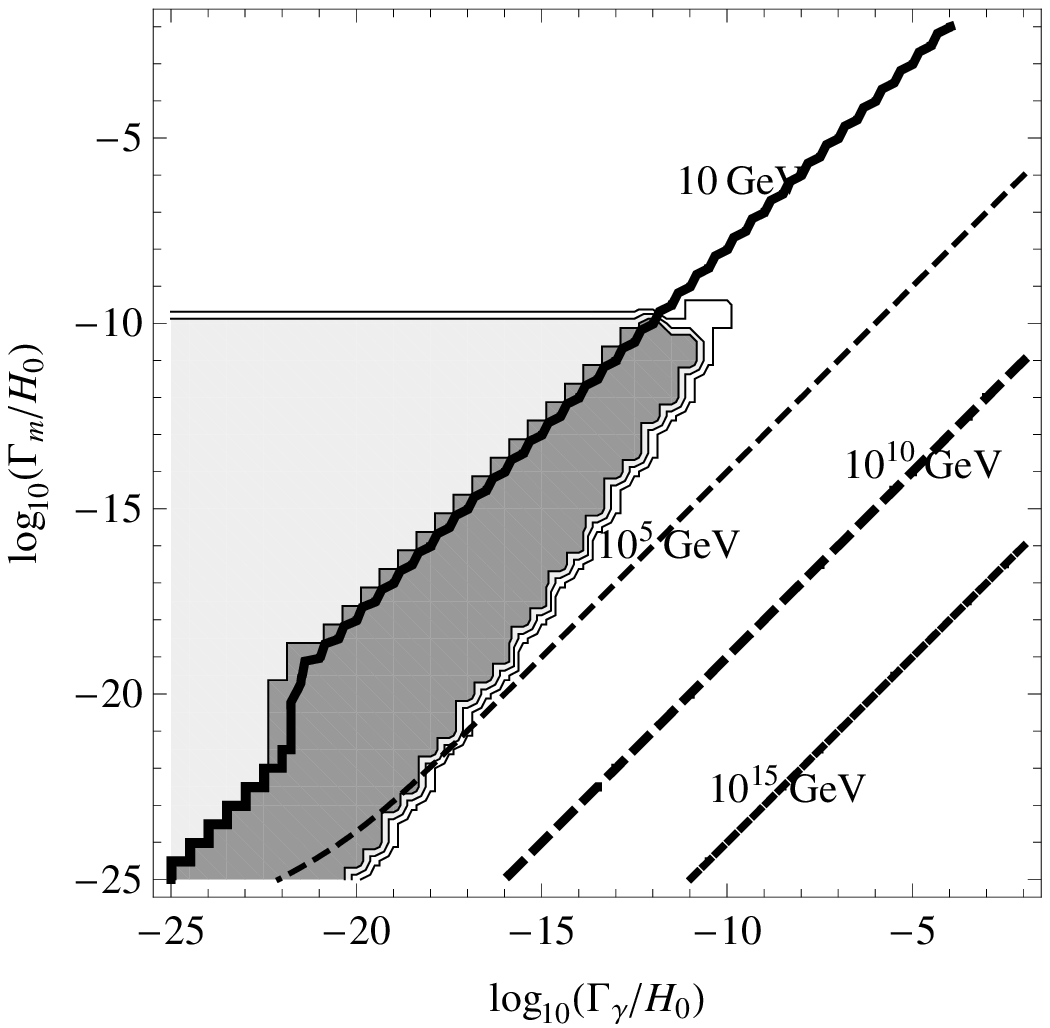}}
\caption{Curvaton-radiation-matter system with different values of $\Gamma_{\gamma}$ and 
$\Gamma_{m}$, when (a) $\Omega_{\sigma0}=10^{-2}$, (b) $\Omega_{\sigma0}=10^{-5}$, (c) 
$\Omega_{\sigma0}=10^{-7}$ and (d) $\Omega_{\sigma0}=10^{-10}$. Thick lines represent the initial 
system temperature in units of GeV.}

\end{figure}

We have rigorously searched for initial values for which the system is physically 
motivated \ie it passes the tests mentioned in the previous section. This case has been 
previously studied in \cite{Gupta:2003jc}.
In figures \ref{fig1a}-\ref{fig1d} these tests can be seen as shades of black and white. A 
system that passes all of the above mentioned tests is labeled by white color whereas the 
opposite case is black. Shades of gray indicate that some but not all of the aforementioned
physical requirements have been fulfilled. Besides these tests we have also included different 
contours of the initial system temperature into the figures.

From the figures, we can read out that smaller values of initial curvaton density in general 
indicate higher initial temperatures, or higher reheat temperatures. The highest reheat temperatures 
correspond to higher values of $\Gamma_{\gamma}$, \ie larger portion of the curvaton decays into
 radiation and 
therefore pushes the time of radiation-matter equality later leading to higher initial 
temperatures. Since the model applies only when the curvaton field is oscillating, this means 
that the temperature during reheating has been even higher. However as can be seen from figures, 
decreasing $\Omega_{\sigma}$ an additional area becomes allowed corresponding to larger values 
of $\Gamma_{m}$. This is a result of the fact that if $\Gamma_{\gamma}$ and $\Gamma_{m}$ are too 
small the curvaton does not decay fast enough, leading to possible issues during nucleosynthesis 
as mentioned above (\eg for $\Omega_{\sigma0}=10^{-10}$, $\Gamma_{\gamma}/H_{0}=10^{-20}$ and 
$\Gamma_{m}/H_{0}=10^{-20}$.) By increasing $\Gamma_{m}$ the curvaton field decays before 
nucleosynthesis and therefore leads to a physically sound system (\eg for 
$\Omega_{\sigma0}=10^{-10}$, $\Gamma_{\gamma}/H_{0}=10^{-20}$ and $\Gamma_{m}/H_{0}=10^{-6}$.)

The figures indicate that if $\Gamma_{\gamma,m}\sim H$, a small initial curvaton density is 
required in order for the system to be physically viable. In addition, we see that in this case 
the reheat temperature is typically quite low. 

\subsection{Time dependent interactions, $f_{\gamma}(N)=1$, $f_{m}(N)=\theta(N-N_{*})$}

\begin{figure}[ht]

\subfigure[]{\label{fig3a}\includegraphics[width=0.45\columnwidth]{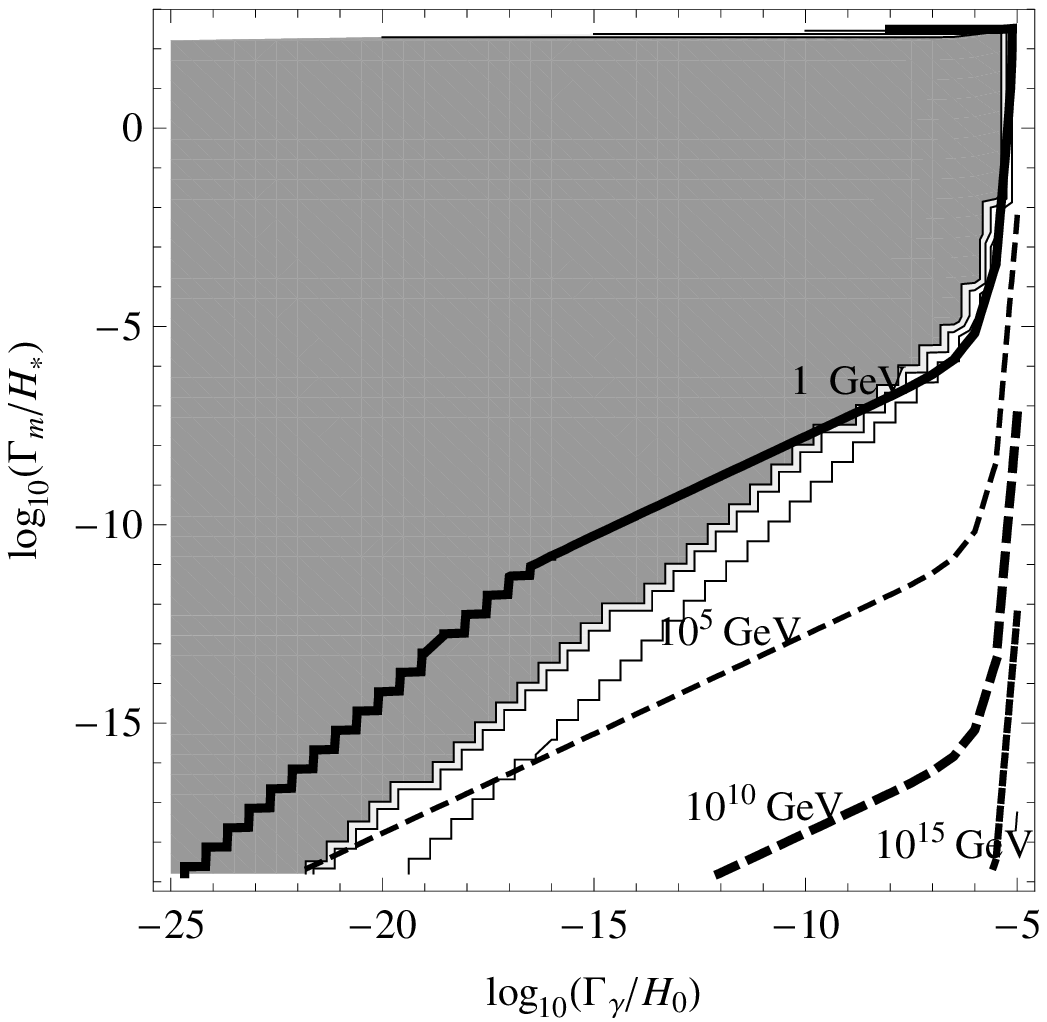}}
\qquad
\subfigure[]{\label{fig3b}\includegraphics[width=0.45\columnwidth]{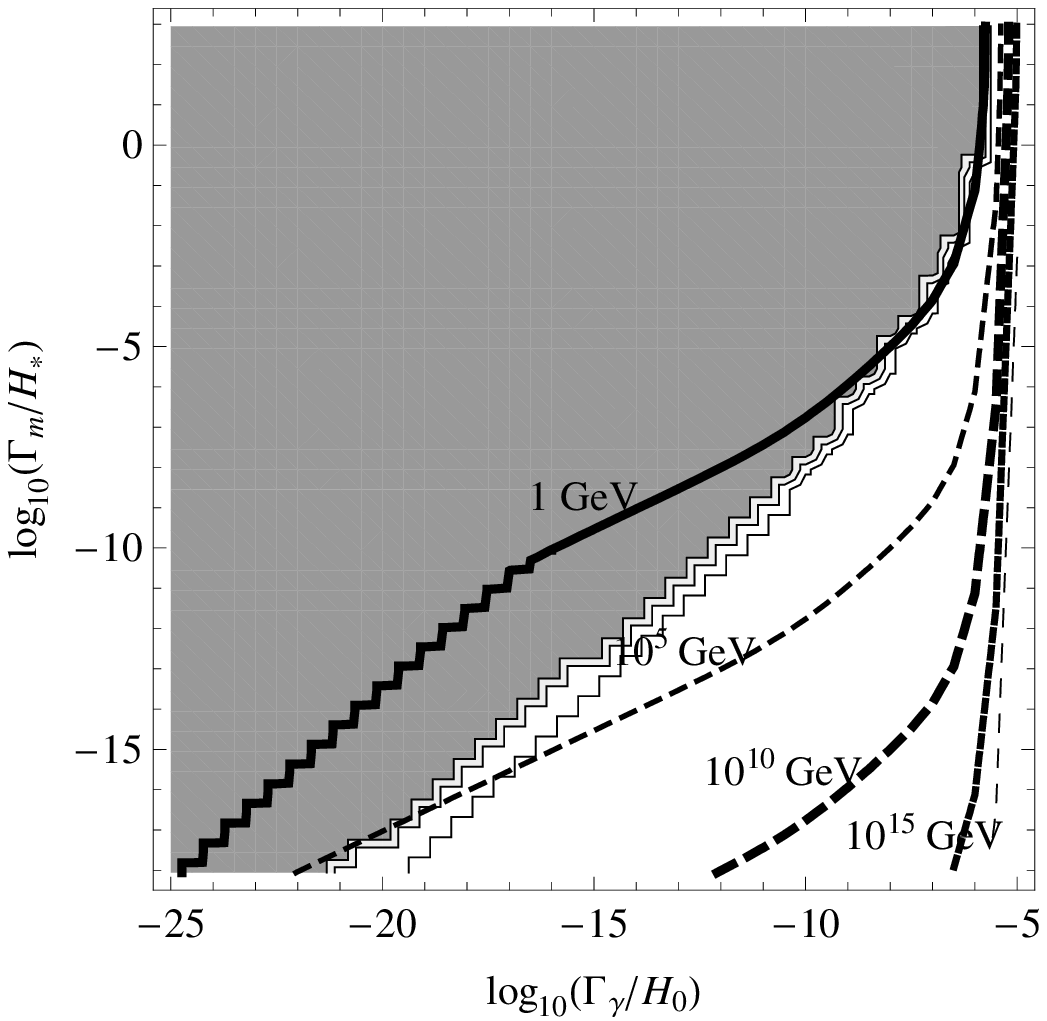}}
\subfigure[]{\label{fig3c}\includegraphics[width=0.45\columnwidth]{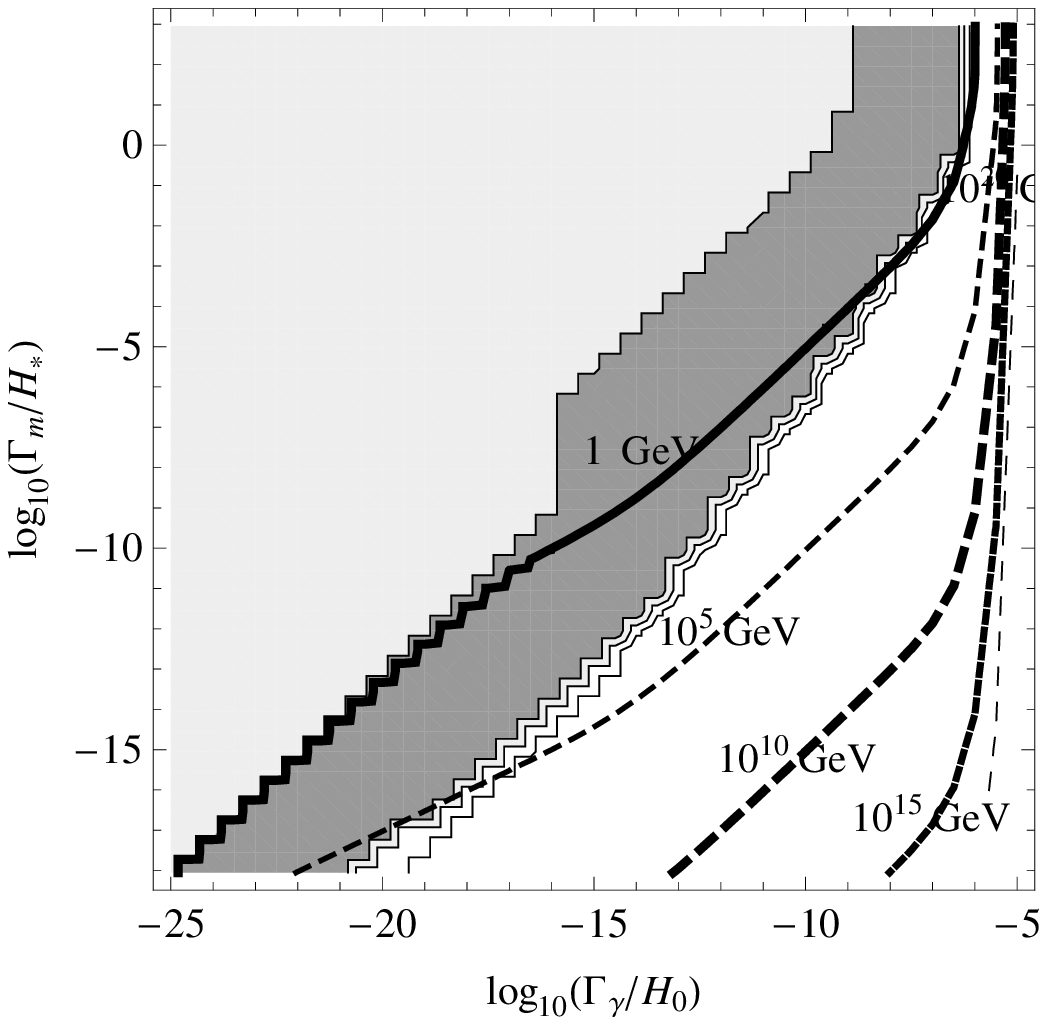}}
\qquad
\subfigure[]{\label{fig3d}\includegraphics[width=0.45\columnwidth]{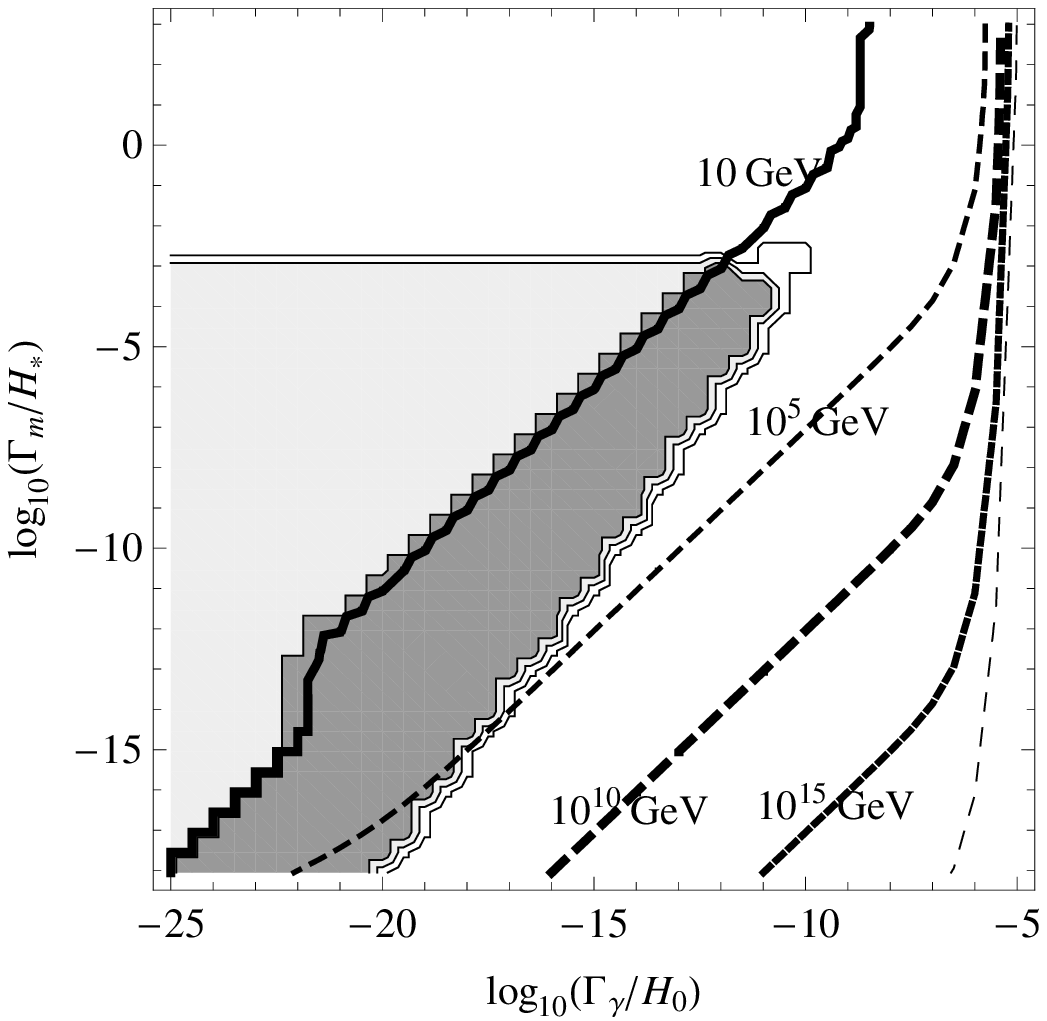}}
\caption{Curvaton-radiation-matter system where matter interaction is turned on at $N=8$ with 
different values of $\Gamma_{\gamma}$ and $\Gamma_{m}$, when (a) $\Omega_{\sigma0}=10^{-2}$, (b) 
$\Omega_{\sigma0}=10^{-5}$, (c) $\Omega_{\sigma0}=10^{-7}$ and (d) $\Omega_{\sigma0}=10^{-10}$. 
Thick lines represent the initial system temperature in units of GeV.}
\end{figure}
 
When the curvaton-matter interaction is turned on at $N_{*}$ the system is driven towards an 
equilibrium of $\zeta_{\sigma}\rightarrow-\phi(N_{*})$. 
This can be seen clearly from the equation of motion of $\zeta_{\sigma}$, \ie 
eq. (\ref{eq:ze-pert}), because at times close to $N_{*}$, terms involving $f'_{m}(N)$ dominate the evolution 
and $\rho_{\sigma}<0$. Thus the time dependent scenario resembles 
the situation we previously studied in \cite{Multamaki:2006}. 
As can be seen from figures \ref{fig3a}-\ref{fig3d}, different initial values lead to 
allowed regions very similar to the previous case. Major alteration comes from the change of 
$H_{0}\rightarrow H(N_{*})<H_{0}$, which shifts different regions upward compared to the time 
independent scenario. Since the studied cases are initially radiation dominated, the Friedmann
equation  $H^2 = \rho_0 \simeq \rho_{\sigma} + \rho_{\gamma} = \rho_{\sigma0}e^{-3N} 
+ \rho_{\gamma0}e^{-4N} \simeq \rho_{\gamma0} e^{-4N}$ allows us to estimate 
$H(N_*)/H_0 \simeq 10^{-7}$ when $N_*=8$. If the initial contribution of curvaton will be larger 
the system will become curvaton dominated earlier and therefore lead to a smaller value of 
$H(N_{*})$.

Another change can be seen in the initial system temperatures which rise steeply when 
$\Gamma_{\gamma}$ increases. Still, higher values of $\Gamma_{\gamma}$ lead to a larger 
contribution of radiation as in the time independent case. However, since now the matter 
interaction is turned on later, the curvaton might decay completely to radiation giving very low 
or no matter contribution at all. This in turn pushes initial temperatures up. Thus
again we see that the initial curvaton density needs to be small. But a difference to 
the continuous-interaction case is that now the reheat temperature can be much higher.

\subsection{Isocurvature}

Isocurvature is defined as the difference between two curvature perturbations 
$S_{m\gamma}=3(\zeta_m-\zeta_{\gamma})$. Since the matter curvature perturbation $\zeta_{m}$ 
is here ultimately always driven to the value $\zeta_{\sigma0}$ \cite{Gupta:2003jc} the amount of 
isocurvature depends only on the final value of $\zeta_{\gamma}$.
The behavior of the radiation perturbation in the three-fluid model has been studied previously 
in \cite{Gupta:2003jc}. They found that if the curvaton begins to dominate before the decay 
epoch almost all of the radiation originates from the curvaton fluid and therefore gives 
$\zeta_{\gamma} \simeq \zeta_{\sigma0}$.

\begin{figure}[tbh!]
\subfigure[]{\label{fig-isoca}\includegraphics[width=0.45\columnwidth]{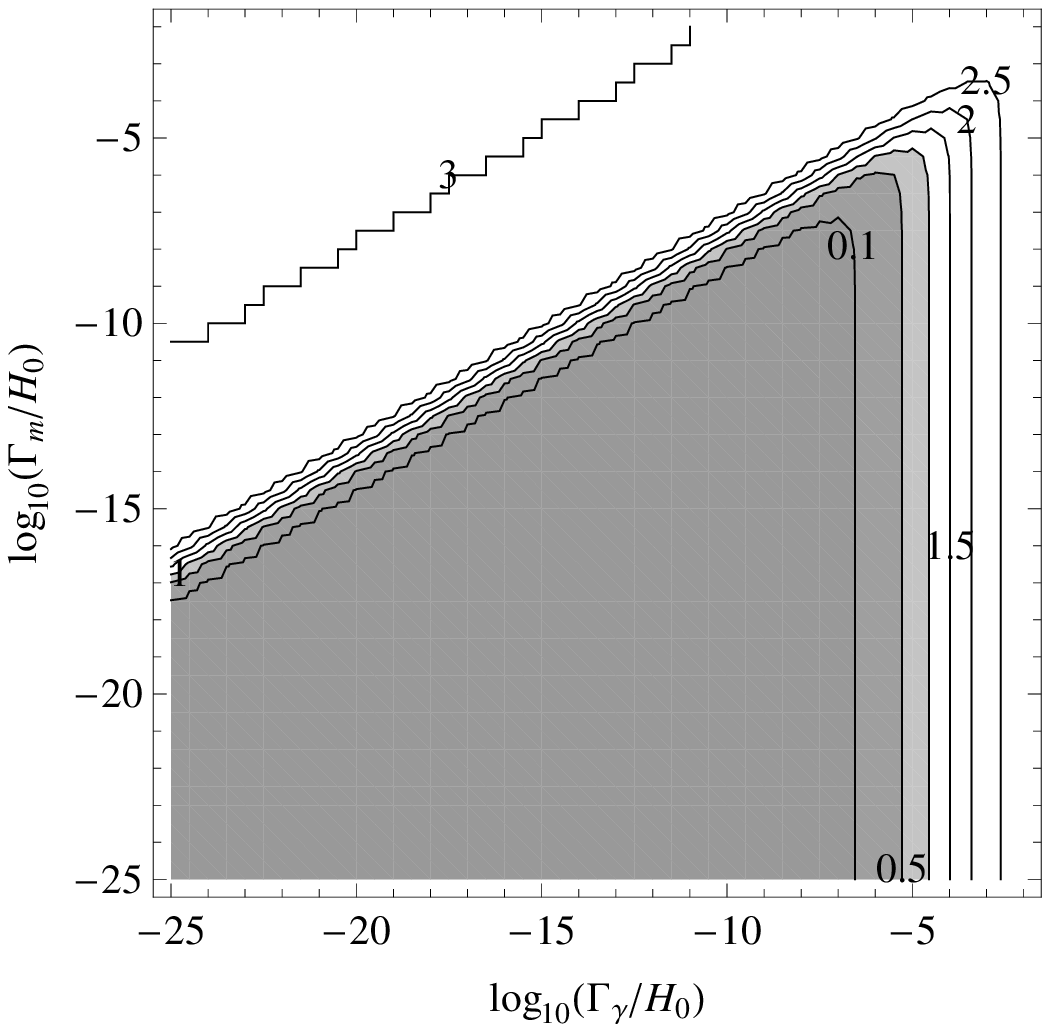}}
\qquad
\subfigure[]{\label{fig-isocb}\includegraphics[width=0.45\columnwidth]{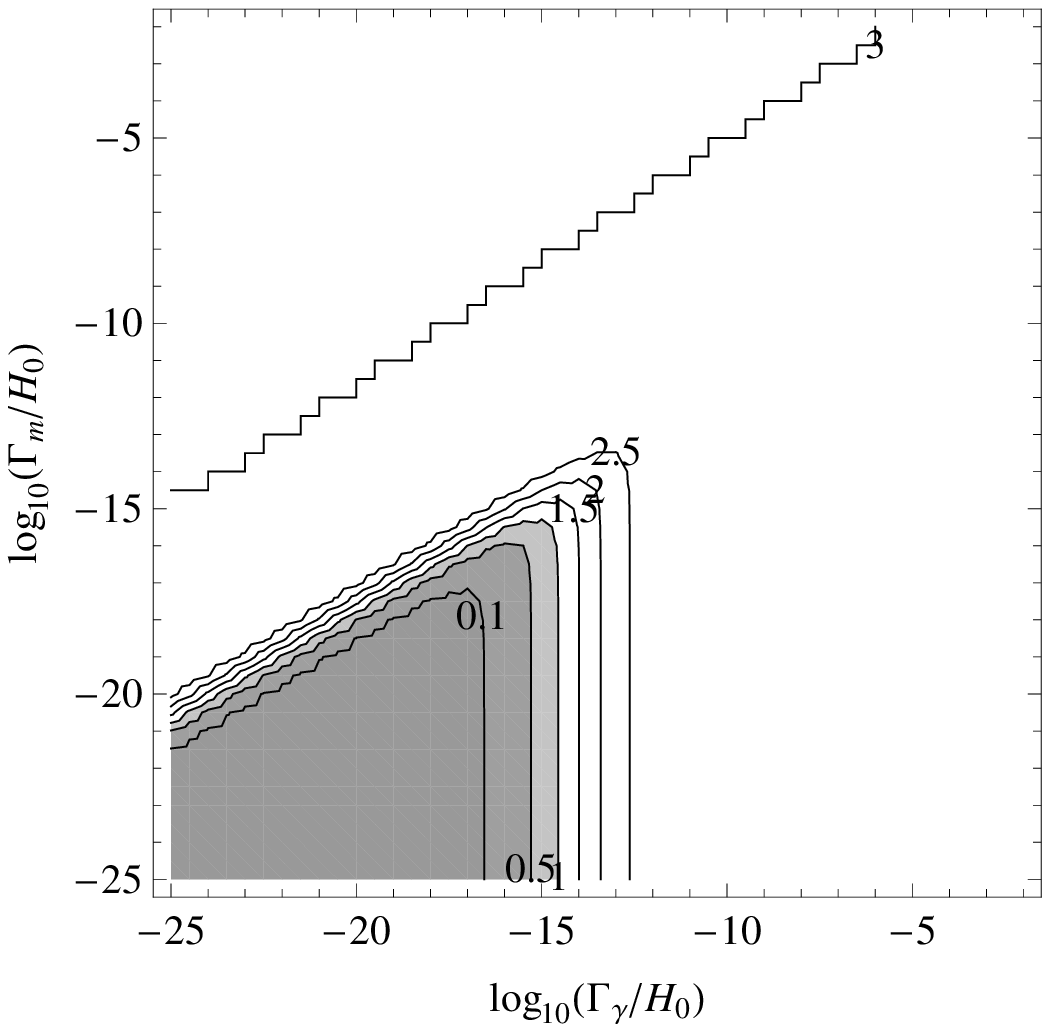}}
\caption{Contours of the isocurvature/adiabatic ratio $S_{m\gamma}/\zeta\Big|_{\textrm{dec}}=
3(\zeta_m-\zeta_{\gamma})/\zeta_m\Big|_{\textrm{dec}}$ at the time of decoupling when (a) 
$\Omega_{\sigma0}=10^{-2}$ and (b) $\Omega_{\sigma0}=10^{-7}$. In the darkest region the system 
is adiabatic whereas the white color corresponds to the largest amount of isocurvature. Larger 
initial contribution of curvaton clearly gives a larger adiabatic region.}
\end{figure}

We have plotted the amount of isocurvature in the three-fluid model in figures \ref{fig-isoca} 
and \ref{fig-isocb}. Our results clearly agree with the reasoning above. If $\Gamma_{m}$ 
or $\Gamma_{\gamma}$ is large enough, the curvaton fluid will decay before it begins to dominate 
the system and hence lead to a non-vanishing final isocurvature. For a smaller initial curvaton, 
it takes even longer for the curvaton fluid to dominate and therefore the system is adiabatic 
only for very small values of $\Gamma_{m}$ and $\Gamma_{\gamma}$. These in turn lead to high 
reheating temperatures,\eg  for initial values $\Omega_{\sigma0}=10^{-10}$, $\Gamma_{\gamma}/H_0
=10^{-26}$ and $\Gamma_{m}/H_0=10^{-34}$ a numerical evaluation gives a reheating temperature 
$10^{12}$ GeV and $S_{m\gamma}/\zeta\Big|_{\textrm{dec}}=0.0006$. The time dependent interaction 
gives very similar results once the scaling $\Gamma_m/H_{0}\rightarrow \Gamma_m/H_{*}$ is taken 
into account.

\section{Discussion and conclusions}

The curvaton model has gained a lot of attention in the recent years mainly because it can make 
the inflation potential look more natural \cite{Lyth:2001nq}.  Since the first model where the 
curvaton decayed only into radiation \cite{Lyth:2001nq,Moroi:2001ct}, a number of other 
possibilities have been explored including a curvaton web model \cite{Linde:2005yw}, the 
possibility of multiple curvaton fields \cite{Assadullahi:2007uw} and different particle models 
such as axions \cite{Chun:2004gx,Dimopoulos:2005bp} just to name a few.

In the present paper we have studied a three-fluid model in which the 
curvaton decays into both radiation and cold dark matter. This has been studied previously also 
in \cite{Gupta:2003jc,Lyth:2002my,Gordon:2002gv}. We have assumed that the initial system has no 
matter content and it is dominated by the radiation which originates from the decay of the 
inflaton field. Because all of the matter content of the universe comes from the curvaton field
we are able to estimate the reheating temperature. This can be additionally used to
constraint the parameter space of the model.

We have systematically scanned the parameter space and identified the regions 
where the model is physically acceptable, \ie when evolution during and after nucleosynthesis is 
standard while requiring that the reheating temperature is not unreasonably high.
We have identified these regions both when the decay 
rates are fixed and when the curvaton starts to decay later. These allowed regions are 
alike once the rescaling $\Gamma/H_0 \rightarrow \Gamma/H(N_*)$ is taken into 
account and the real difference appears in the initial system temperature.

We find that if the decay rates are comparable to the Hubble rate, a small initial 
curvaton density is required. Otherwise one needs to fine-tune the decay rates to be much 
smaller than $H$ at the time of decay. In the continuous interaction case, requiring 
$\Gamma_i\sim H$ leads to a low reheat temperature, but this can be avoided when the matter 
interaction is delayed. 

If the initial curvaton density is large, the final state is naturally adiabatic assuming
that the system is otherwise physically acceptable. Note however, that this in turn requires 
fine-tuning in the decay rates. If $\Gamma_i\sim H$, we find that the final state generally 
contains a large isocurvature component.

We have also we studied non-gaussianity in the framework of the three-fluid models. We find that 
in the region where the first-order perturbation theory can be applied, the three-fluid model 
gives no limits on the $f_{NL}$ parameter. This is the result of a conserved curvature 
perturbation $\zeta_c$, which carries the initial curvaton perturbation $\zeta_{\sigma}$ into 
the matter perturbation $\zeta_m$ \cite{Gupta:2003jc}. Our results differ from the previous 
results \cite{Lyth:2002my,Gupta:2003jc} mainly because our $f_{NL}$ is evaluated at the time of 
last scattering and not at nucleosynthesis. This allows the non-gaussianity to be more easily 
compared to the observational Sachs-Wolfe effect and we do not have to use a radiation transfer 
function. In order to calculate the observational non-gaussianity, second-order perturbation 
theory needs to be applied \cite{Bartolo:2004if}. This is however beyond the scope of this 
article and will be the focus of a follow-up paper.

\subsection*{Acknowledgments}

This project has been partly funded by the Academy of Finland project no. 8111953.
TM and JS are supported by the Academy of Finland.



\end{document}